\documentclass[journal]{IEEEtran}
\usepackage{comment}
\usepackage{hyperref} 
\usepackage{titlesec}
\usepackage{tabularx, booktabs}
\usepackage{xcolor}
\usepackage{ifpdf}
\usepackage{cite}

\ifCLASSINFOpdf
 \usepackage[pdftex]{graphicx}
\else
 \usepackage[dvips]{graphicx}
\fi

\usepackage{amsmath}
\usepackage{algorithmic}
\usepackage{array}
\ifCLASSOPTIONcompsoc
  \usepackage[caption=false,font=normalsize,labelfont=sf,textfont=sf]{subfig}
\else
  \usepackage[caption=false,font=footnotesize]{subfig}
\fi
 \usepackage{dblfloatfix}
 
\usepackage{url}
\begin{document}
\title{Machine Learning Applications in Cascading Failure Analysis in Power Systems: A Review}
\author{Naeem~Md~Sami,~\IEEEmembership{Graduate Student Member,~IEEE}, and Mia~Naeini,~\IEEEmembership{Senior Member,~IEEE}

\textit{Department of Electrical Engineering, University of South Florida, Tampa, FL 33620, USA}

naeemmdsami@usf.edu, mnaeini@usf.edu \vspace{-2em}}

\maketitle

\begin{abstract}
Cascading failures pose a significant threat to power grids and have garnered considerable research interest in the power system domain. The inherent uncertainty and severe impact associated with cascading failures have raised concerns, prompting the development of various techniques to study these complex phenomena. In recent years, advancements in monitoring technologies and the availability of large volumes of data from power systems, coupled with the emergence of intelligent algorithms, have made machine learning (ML) techniques increasingly attractive for addressing cascading failure problems. This survey provides a comprehensive overview of ML-based techniques for analyzing cascading failures in power systems. The survey categorizes these techniques based on the evolutionary phases of the cascade process in power systems, as well as studies focusing on cascade resiliency before the occurrence of cascades and problems related to cascades after their termination. By organizing and presenting these works into relevant categories, this survey aims to offer insights and a systematic understanding of ML's role in mitigating cascading failures in power systems.
\end{abstract}


\IEEEpeerreviewmaketitle

\vspace{-0.3cm}
\section{Introduction}
\label{sec:sec1}

\subsection{Overview and Significance}
\label{sec:sec1A}
Despite the advancement of modern power grids, which are equipped with increasingly intelligent monitoring, control, and communication systems, historical data indicate that they remain susceptible to various cyber and physical stresses. The reliability-threatening stressors can affect all layers of these systems; however, stresses on transmission networks can
have widespread and devastating effects such as large blackouts \cite{def2}. 
While the N-1 security criterion has traditionally been used to assess the reliability of power systems, it is important to recognize that power grids can still be susceptible to physical stresses, such as multiple contingencies arising from natural disasters or deliberate sabotage. Such failures along with the lack of timely and effective control actions can trigger a sequence of interdependent component failures in power systems, called {\it cascading failures}, leading to large blackouts. 

Cascading failures and diffusion phenomena manifest in a multitude of real-life complex systems, exhibiting diverse forms and scales. Examples include the spread of information in social networks, the propagation of computer viruses, and the transmission of infectious diseases in communities \cite{def7, def8}. 
Extensive research has been conducted to comprehend these processes, resulting in the development of various models as well as techniques for their control. While studies have demonstrated similarities in these processes across different systems, the contributing factors and underlying interaction mechanisms in cascades differ among systems. Generally, cascading failures and spreads are influenced by complex interactions of the large number of components within the system, which are further impacted by various attributes and characteristics unique to each system. 
In the context of power grids, analysis of historical data on cascading failures and blackouts, such as the 2003 Northeast blackout \cite{refe4} and the 2011 Southwest blackout \cite{def10}, highlights that the cascade process is not solely determined by physical component failures or associated physics-based interactions. Other factors, including the system's operating settings \cite{OperatingSetting}, cyber vulnerabilities \cite{PowerCommDep} (e.g., computer server failures and communication issues), and human factors \cite{def11, def12}, also play a role in influencing the cascade process.
Moreover, it has been discussed that failure of state estimator models and lack of situational awareness were significant contributors to these events and a timely reaction (such as load curtailment or islanding) could have significantly reduced the impact of these events \cite{def13}. 

Considering the complexity of these phenomenon and challenges in controlling them, a large body of work has been formed in understanding cascading failures and mitigating their effects in power grids.   
Particularly, cascading failures in power grids have been studies using power physics-based approaches \cite{m7, m21, m8}, simulation-based techniques \cite{refe6, refer4, m10}, probabilistic models \cite{v2, m13,m38}, and graph-based modeling and analyses \cite{refe13, refe29}.
Despite all the studies and developed techniques, due to the large size and geographical scale, complex and at-distance underlying interactions among the components, and new attributes and dynamics of modern power grids (for instance,  deployment of stochastic renewable resources), cascading failures have remained, although not very common, but a complex and costly threat to these systems. 
With the advances in monitoring and computational technologies in power systems and the availability of large volume of energy data generated from the deployed sensors and metering devices (such as phasor measurement units--PMUs) as well as datasets of historical and simulated cascading failures, Machine Learning (ML) based techniques have been recently adopted in studies of cascading failures in order to complement the traditional techniques and to shed light on embedded patterns and signatures of stresses and cascades in power systems for various descriptive, predictive and prescriptive analyses. 

\vspace{-0.3cm}
\subsection{Motivation and Contribution}
\label{sec:sec1B}
Several surveys have reviewed various aspects of cascading failure studies, including modeling, analysis, benchmarking, and validation of analysis tools \cite{review1, review2, review3, review4, review5,review8}. As the use of intelligent algorithms in cascade analysis is on the rise, it is important for the scientific community to have a birds-eye view of the possible routes in addressing the complex problems related to cascading failures in power systems using such techniques.
Furthermore, as power systems become more complex and exhibit stochastic behavior due to their expanding scale and the integration of renewable resources, and as the deployment of measurement devices generates vast volumes of data, the utilization of data-centric, ML-based techniques becomes crucial to support monitoring and decision-making processes before, during, and after cascading failures.
As such, the objective of this survey is to examine the utilization of data analytics and  ML techniques in the analysis of cascading failures within power systems, while also addressing the existing gaps and unresolved issues pertaining to cascading failures in power systems that can be potentially addressed using these techniques.
In addition to their application in cascading failures, ML techniques have been widely employed in various other domains within power systems. These applications include system monitoring, state estimation, fault diagnosis and prediction, load forecasting, power system security, energy management, and optimization. Example surveys on such ML applications include \cite{def22, def23}.
 
One of the primary contributions of this paper lies in its systematic organization of the review and the discussions, guided by different phases of cascading failures. Historical data and simulations of cascading failures have revealed that the time evolution of cascades exhibits three phases \cite{refe4, def10, def11}. 
The first phase, known as the {\it precursor phase}, is characterized by a slow progression of failures. During this phase, control actions such as dispatching, load shedding, and intentional/controlled islanding can effectively mitigate the impact of disturbances. 
The second phase is the {\it escalation phase}, in which failures occur rapidly and preventing blackouts becomes significantly more challenging. 
The third and final phase is the cascade {\it phase-out}, where the rate of failures slows down as a significant number of components have already failed. 
In addition to reviewing works focused on various analyses during the cascade process, this paper also examines ML techniques developed for cascade resiliency before the cascade occurs, as well as problems related to cascades after their termination.
The categories of the work related to these phases are introduced next. 

\begin{itemize}
    \item {\it Normal Phase Category:} This category includes works that address problems related to the cascade under the normal operating conditions of the system when there is no disturbances in the system. 
    The works in this category are mainly focused on vulnerability analysis with respect to cascading failures, network hardening efforts toward cascading failures and cascade modeling and simulations.
    \item  {\it Cascade Precursor Phase Category:} 
    This category encompasses studies that tackle issues pertaining to cascades after the system has undergone disturbances and the system's state regarding the risk of cascades.
    The works in this category are mainly focused on performing initial fault analysis, various predictive analysis and deciding corrective measures to prevent a large cascade. Examples of predictive analysis in this phase include predicting the size of the cascade, predicting the evolution and temporal aspects of the cascade, and predicting the path of the cascade and areas at risk.  Examples of corrective measures to be identified include dispatching, load shedding, and islanding.
    \item {\it Cascade Escalation Phase Category:} This category has been designated as a temporary classification for problems associated with the escalation phase in the cascade process within power systems. As previously mentioned, during this phase, failures rapidly propagate throughout the system, leading to an inevitable large-scale blackout. Once the cascade reaches this phase, there are limited options for intervention. Consequently, there is a scarcity of studies specifically focusing on this particular phase.
    \item {\it Post Cascade Termination Phase Category:} 
    This category comprises works that concentrate on problems after the cascade process has ended and a blackout has already taken place. The studies in this category primarily focus on cascade fault recovery, service restoration, and conducting root cause analysis of the cascade that occurred.
\end{itemize}

\begin{figure*}
\centering
\includegraphics[width=15cm]{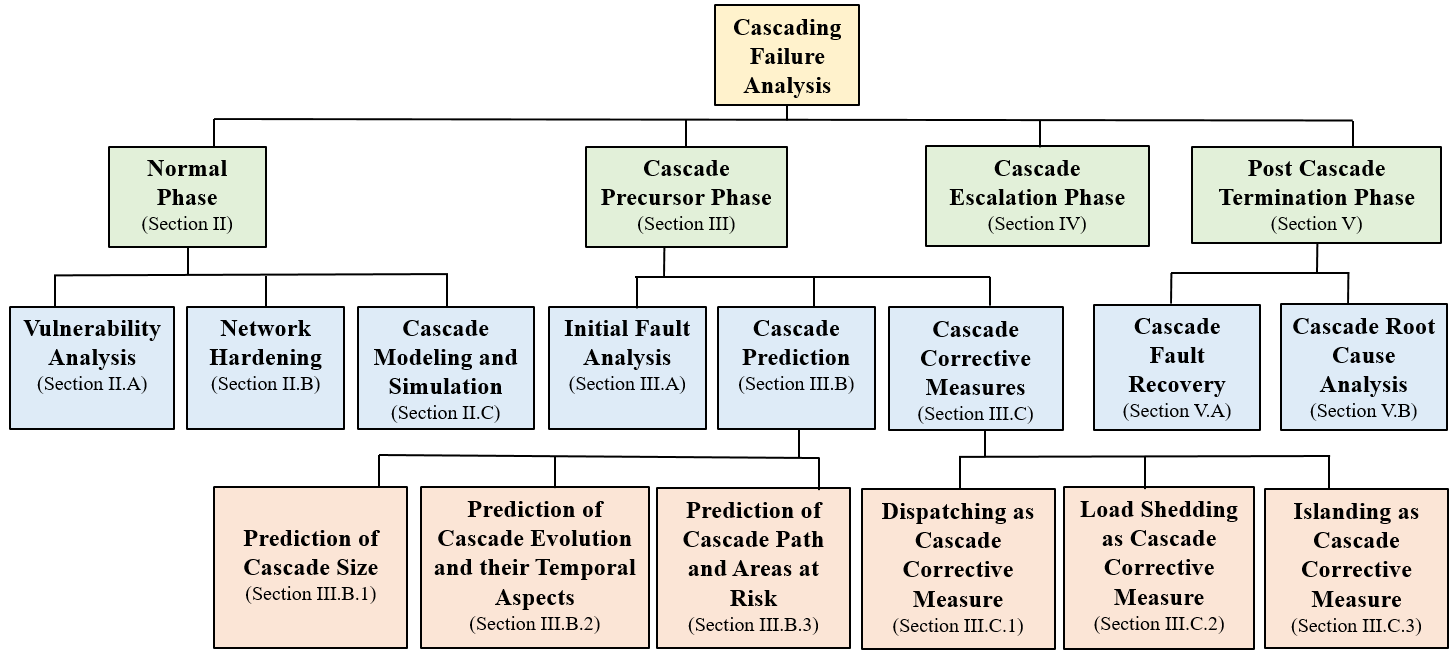}
\caption{Taxonomy of cascading failure analysis supported by ML methods based on different phases before, during and after cascading failures. The section numbers associated with the categories and sub-categories are marked in each element.}
\label{fig:fig1}
\end{figure*}

These categories of analyses and studies and their association with different phases of cascade are shown in Fig.~\ref{fig:fig1}.

In this survey, a comprehensive overview of the state-of-the-art applications of ML in analyzing cascading failures in power systems is presented by reviewing a large number of journal and conference articles from reputable databases. While it is possible to further categorize these publications based on cascade models and types of algorithms, the categorization based on cascade phases proves to be more useful in discussing the existing gaps in supporting critical functions for cascade mitigation and resiliency. The applications of ML in the cascade analysis in these categories is summarized in Table~\ref{table:table1}.

\section{ML-based Cascade Analysis in Normal Phase}
\label{sec:sec2}
This category reviews research that focuses on using ML for addressing issues related to cascades within the normal operating conditions of the power system, where no disturbances or disruptions are present. These works are further classified into vulnerability analysis, network hardening, and cascade modeling and simulations.
\vspace{-0.2cm}
\subsection{Vulnerability Analysis}
\label{sec:sec2A}
Vulnerability analysis in power grids involves identifying critical components in the system that carry a higher risk of failure or that their failures can lead to higher reliability concerns and larger service disruptions. The focus of the review in this section is particularly on the works for identifying vulnerable components that can lead to cascading failures in power systems. Identifying such components before failure is important to eradicate potential risks of system and service impairments and to implement protection and mitigation mechanisms.

Identifying combination of components that their simultaneous fault can trigger cascading failures \cite{b6, b3, tox6, tox4} as well as identifying sequence of failures or attacks that can cause cascades \cite{b7, refer77}, both in power systems and interdependent power and other critical infrastructures, such as communication systems \cite{refer64, refer70} and gas systems \cite{refe6}, have been considered in the studies of cascading failure vulnerabilities. Moreover, vulnerability of cyber components related to cascading failures have also been studied in the literature \cite{refer57, refer63, refer68}. Another closely related problem to vulnerability analysis is identifying critical components, the protection of which can reduce the risk of cascades \cite{refer58, b4, tox15, refer70}. Although such components may not be the vulnerable ones triggering the cascade, their vulnerability can fuel the cascade process. Overall, various forms of cascade vulnerability analysis have been studied in the literature using simulation-based approaches \cite{refer77, b4, tox15, refe6}, graph-theoretic approaches \cite{refer55, refer59, refer60}, game-theoretic approaches \cite{refer68, refer72, refer73}, and optimization techniques \cite{refer67, refer70, tox4}.  Next, the ML-based techniques applied to these problems are briefly reviewed. 

The research on the application of ML to vulnerability analysis has been classified into three categories, based on their modeling approach. These categories are search, classification, and regression-based approaches.

{\bf Search Approach:} The vulnerability analysis can be modeled as a search problem in power grids. Searching power grids for vulnerable components that can trigger or fuel cascading failures is a daunting task due to the large number of combinations of failures that can be considered and can occur in cascades. The goal of ML-based approaches is to support these analyses for a more accurate and efficient search process over power systems. 
A large number of data-driven and intelligent algorithms for efficient search of vulnerable components and the most impactful attack sequences including greedy search algorithms \cite{tox6}, particle swarm optimization methods \cite{b7, refer79}, genetic algorithms \cite{refer80, refer81, refer82} and random chemistry techniques \cite{refer67, refer31}  as well as reinforcement learning \cite{refer77, b4} and deep learning approach \cite{b5, c10} have been developed, creating a solid scope for the application of similar algorithms for efficient search over the power grids.

Although the mentioned heuristic algorithms perform direct search over the grid topology, reinforcement and deep learning-based search techniques took data-centric approaches for pattern learning. The work in \cite{b4} suggests a temporal difference reinforcement learning mechanism to learn the relation between faults and load loss to identify the fault chain that leads to the largest load loss. The authors in \cite{refer77} use a Q-learning technique to search for the critical lines when the grid is under sequential attack. When the grid topology changes under such an attack, the technique learns the sequence of the lines that will lead the grid to a specific blackout size. In \cite{b5}, a search framework with the aid of a graph convolutional network (GCN) is developed to identify critical cascading failures. The GCN model guides the search by classifying between normal and load shedding outcomes based on the state of the system defined through input features including the state of the lines. 
The work in \cite{c10} develops a deep convolutional neural network to classify the risk level of transmission lines based on their topological and operational characteristics, and the depth first search algorithm is used to identify the critical lines that can trigger cascading outages. 

{\bf Classification Approach:} Another group of works models the cascade vulnerability analysis as a classification problem to classify the components of the system as, for example, robust, normal, and vulnerable groups. For instance, the work presented in \cite{tox15} utilizes features such as the centrality of the buses (i.e., betweenness centrality), and power flow information for vulnerability classification of buses using XGboost, which is an optimized distributed gradient boosting library. The latter approach has been compared with classification with logistic regression, support vector machine, and k-nearest neighbors in \cite{tox15}.
In \cite{refe6}, a steady-state energy flow model for a combined power-gas system has been considered and a random forest hybrid classification-regression is formulated to classify the vulnerable power and gas components. Specifically, the regression model is used to predict the vulnerability metric for each of the components enabling the classification of the vulnerable components.

{\bf Regression Approach:} 
Regression-based analysis of vulnerable components focuses on learning vulnerability metrics for the components of the system. Similar to the work in \cite{refe6}, which used a regression model to predict the vulnerability metric for each of the components, the work in \cite{c1} uses a graph neural network-based prediction of avalanche centrality, which is the measure of the impact of a node on the avalanche dynamics of a Motter-Lai cascading failure model. A probabilistic risk index based on load shedding, voltage violation, and hidden failure is predicted in \cite{refer76} using decorrelated neural network ensemble to understand $N-k$ contingency analysis considering potentially cascade inducing outages.

\vspace{-0.3cm}
\subsection{Network Hardening}
\label{sec:sec2B}
The primary objective of network hardening in power systems is to enhance the resilience of the system 
against faults, attacks, and stresses that have the potential to trigger cascading failures.
In this field, existing research includes various approaches to network hardening, some of which involve strengthening the system's structure through methods such as increasing redundancy (e.g., adding extra components and connections)\cite{a3} or resilient structure design \cite{refer54, a5}, and improving the protective and preventive maintenance of vulnerable and critical components in the cascade process \cite{b3, a4}.
It is important to note that not all network hardening studies solely focus on cascading failures and they may have a broader scope that includes hardening efforts for the distribution systems \cite{N9,b10}.
Additionally, the literature on this topic includes investigations into severe weather impact analysis \cite{N5, N6} and smart vegetation management \cite{N7, refer45} as means to harden grid components.
The network hardening problem has been addressed through different approaches including optimization \cite{refere13,a1,refer54, b8, b11}, network theoretic approaches \cite{a2,b3,refer54, a4}, and game-theoretic approaches \cite{refere14, refere15}. 
Nonetheless, as far as the authors are aware, the utilization of ML in the transmission network hardening against cascading failures remains an area to be explored.

\vspace{-0.3cm}
\subsection{Cascade Modeling and Simulations}
\label{sec:sec2C}
Cascading failures, which are infrequent occurrences resulting from rare interactions in large-scale power transmission networks, present a challenge when it comes to modeling and simulating them with the necessary level of details for analysis. To address this, several cascade models and simulation platforms have been developed over the years \cite{m37, review3, m41, review7}. Furthermore, numerous studies have focused on simulating various events like cyber attacks, natural disasters, and physical damages that can trigger cascading failures. This section provides a comprehensive review of ML-based techniques utilized in the modeling and simulation of cascading failures. The review is categorized into three main sections:

{\bf ML Application in Power Flow and Failure Propagation Calculations:} Some research in the field of modeling and simulating cascading failures has primarily focused on efficient and rapid power flow calculations, as well as simulations of failure propagation, for example, based on line overloading mechanisms.
For instance, in the work referenced as \cite{refer41}, a physics-informed cascade model is developed using a graph neural network. This model enables faster and more accurate calculation of the power flow by predicting power flow values based on the dynamic equations of the system. Specifically, for cascading failures due to line overloading, the proposed graph neural network performs regression to predict the power flow values and compare the results against AC power flow values using a physics-based loss function. A pre-training and fine-tuning method based on transfer learning is embedded within the model to avoid retraining the model when topological changes occur due to the failure. 
In another study presented in \cite{else5}, an ML approach is adopted to predict power flow values for all branches after each step of cascading failures. This is achieved by utilizing an artificial neural network in an iterative manner, where the output is fed back as input, enhancing the simulation of cascade evolution. Furthermore, in the research outlined in \cite{b18}, a time series interaction model is learned using logistic regression to determine the interaction matrix for changes in line states. The obtained results are then subjected to a binary decision-making process to determine the line status at subsequent time steps to model failure propagation.

{\bf ML Application in Modeling and Characterizing Interactions in the Cascade Process:} 
Comprehending the interactions among system components during the cascade process is crucial for understanding the cascade behavior within the system.
Interactions among system components during cascading failures have been studied and modeled using various forms of interaction graphs \cite{refe29}. To enhance the understanding and modeling of cascading failures, ML techniques have been employed to capture the structures and patterns of interactions among components in power systems' cascades.
For instance, in the study presented in \cite{tox1}, a deep convolutional generative adversarial network is used to learn the failure interaction matrix at each step of the cascading failures. It is discussed that the predicted matrix can either help recovering missing data due to lack of information during cascading failures, or can help discovering new interactions that can provide information about interactions in the next steps.  In another instance, detailed in \cite{refer49}, a spatio-temporal GCN model is developed to learn the importance matrix to reveal power system interconnections for cascade predictions. This model captures the spatial and temporal aspects of the system to determine the significance of different connections. Furthermore, in the research presented in \cite{refer86}, a logistic regression model is designed to learn indirect interactions between the states of lines in order to model failure propagation trajectories after the initial failures. 

{\bf ML Application in Modeling Triggers of Cascading Failures:}
The design of failure and attack models, as well as the identification of the most impactful triggers for cascading failures, is a key problem in cascade modeling and simulation. This problem is closely related to vulnerability analysis discussed in Section~\ref{sec:sec2A}. Modeling such triggers for generating cascading failure has been explored using different approaches such as physics-based \cite{refer90} and game-theoretical techniques \cite{refer72}. ML has also been adopted in this problem. For instance, in the study referenced as \cite{refer33}, reinforcement learning techniques, specifically double Q-learning, are employed to model sequential attacks that have a significant impact. 
The work proposes an attacking scheme that determines the minimum number of attacks needed to cause large cascades, taking into account factors such as line tolerance, the probability of line disconnection, and hidden failures.
To enhance the efficiency of the search process, the approach described in \cite{refer43} utilizes candidate pool-based Q-learning to shrink the search space by focusing on the nodes with the highest loads and degrees. 
\vspace{-0.2cm}

\section{ML-based Cascade Analysis in Precursor Phase}
\label{sec:sec3}
This category includes studies that focus on addressing challenges related to cascades occurring after the system has undergone disturbances. This phase of the cascade is one of the most critical stages to react to the process, prompting extensive investigations and studies to support crucial functions during this period. The aim of these efforts is to enhance the understanding of cascade risks and facilitate informed decision-making regarding corrective actions. Studies falling within this category can be further classified into three classes: initial fault analysis, cascade prediction, and cascade corrective measures.

\vspace{-0.3cm}
\subsection{Initial Fault Analysis}
\label{sec:sec3A}
After the occurrence of initial failures in a power system, it becomes crucial to evaluate the system's state and identify potential risks of cascading failures. Analyzing the impact of initial stresses on the cascade process has been studied using a probabilistic method in \cite{ProbInitailFail} and using a physics-based approach in \cite{refer91}. Examples of ML approaches to stability assessment \cite{ref4} include transient stability assessment \cite{N12,N13,N14, N15}, and the acceleration of N-1 contingency
screening \cite{refer74}. Another related problem is identifying the risk of cascade after the initial failures. For instance, research presented in \cite{refer38} proposes a combined GCN and long short-term memory model to approximate risk parameters as regression targets to evaluate the risk of cascading failures in real time after the initial failures. The work presented in \cite{refer87} proposes using feed-forward neural networks and graph neural networks to perform binary classification of a given graph into safe or unsafe to estimate the risk of cascading failure after the initial failure. Classification of the system state datasets enable proactive early warning for cascades. For instance, the work in \cite{v2} suggests a Support Vector Machine (SVM)-based classification method to classify a given power loading level into normal or potential blackout cases.
\vspace{-0.3cm}

\subsection{Cascade Prediction}
\label{sec:sec3B}
When the power system experiences initial disturbances and risk of cascading failures is assumed, prediction of cascade attributes can facilitate the understanding of the state of the system and identifying potential mitigation strategies to subdue the failures. Predictive analysis of the cascade in power systems involves prediction of cascade size, cascade evolution and its temporal aspects, and propagation path and regions. Although the utilization of ML for outage prediction and power system resilience has been reviewed in \cite{ref4,b20}, respectively, to the best of the authors knowledge, there is no review of data-driven and ML applications in predictive analytics related to cascade attributes in power systems.
This section is focused on reviewing such literature.

\subsubsection{Cascade Size Prediction}
Once the risk of a cascade has been identified, a key challenge is to predict the potential size or magnitude of the cascade. This problem can be approached by characterizing the distribution of cascade sizes as a function of the number of component failures (e.g., transmission lines or generators), the amount of load shed, or the number of affected customers.
The study of cascade size distribution has been extensively explored in the context of power systems, taking into account both historical and simulation data. Researchers have investigated the general form of the distribution of cascading failures and have observed its heavy-tailed power-law nature \cite{Naeini1, m38, m21, b21, igraph2, refe34, refe28}. Similarly, the study of interdependent power and communication systems has also contributed to the understanding of cascade size distributions in interdependent systems \cite{m42, refe13}. Characterizing the cascade size distribution unique to each state of the system and the initial disturbances is one of the focuses in this category of problems. 
In a study presented in \cite{refer29}, the cascade size distribution is examined based on the location of initial failures in various communities identified within the power system. A Markov chain model is utilized to analyze and understand the cascade size distribution in this context.
The cascade size distribution and the effects of influential component failures on cascade size distribution are studied using an influence-based model in \cite{refe36}. 
The prediction of cascade sizes allows operators and decision-makers to assess the severity of the system's state and initial disturbances. This information plays a crucial role in making informed decisions to mitigate the cascade and improve the overall resilience of the system.

The majority of the work in characterizing the probability distribution of cascade sizes 
primarily relies on data-driven and statistical methods, while the direct application of ML to this problem remains limited.
From the available works, the work in \cite{refe13}, applies linear and polynomial regression, decision tree, and deep neural network to predict the total number of components that failed after the initial faults while considering the topological features such as degree and betweenness among the nodes. In \cite{igraph2}, the interaction among the lines of the grid is learned by the expectation minimization algorithm to characterize the probability of small, medium, and large cascade sizes for the number of line failures. In \cite{refe34}, the size of the cascade is predicted through a classification problem with three classes of cascades including no cascade, small, and large cascades using different ML algorithms including logistic regression, k-nearest neighbor, decision tree, random forest, SVM, and Adaboost.

\subsubsection{Prediction of Cascade Evolution and Temporal Aspects} 
As mentioned in Section~\ref{sec:sec1B}, the analysis of historical data indicates that the propagation of cascading failures can be categorized into distinct phases over time. Understanding the temporal aspects and the evolution process of cascades is crucial for operators to recognize the temporal division between these phases and estimate the remaining time available to respond to the situation effectively. This knowledge becomes particularly valuable in assessing the urgency of the situation and taking appropriate actions before the cascade transitions into the escalation phase. 

Despite the significance of understanding and predicting the temporal aspects and evolution of cascading failures, there is a limited body of existing research in this domain. For instance, the work in \cite{m38} uses a data-driven Markov model to characterize the evolution of the blackout probability over time.
In another work, \cite{p5}, ML techniques are applied to predict the onset time of the cascading acceleration phase. The problem is formulated as a multi-class classification task, where a neural network is employed to classify a given scenario into urgent, relatively urgent, or non-urgent categories based on the predicted onset time. While this work focuses on categorizing the temporal urgency of cascade scenarios, the detailed exploration of onset time itself and the broader evolution aspects of cascades remain relatively unexplored.
The limited existing work highlights the need for further research and development in understanding and predicting the temporal dynamics and evolution process of cascading failures.

\subsubsection{Predicting Cascade Path and Areas at Risk} 
Given the non-local nature of cascade propagation and the complex interactions among components in power systems \cite{refe28}, accurately predicting the specific areas and components that will be affected by the cascade process poses a significant challenge. Consequently, research in this category is dedicated to predicting the components or regions in the system that are at a higher risk of failure following the initiation of a cascade. Within this area, there are various subproblems that researchers address. Some focus on predicting the next component failure, while others aim to predict the next $k$ failures. Additionally, some studies seek to predict the complete path of the cascade, including all subsequent failures. 

For instance, in the work presented in \cite{refer29}, the focus is on identifying the locality of cascading failures in relation to the underlying communities within data-driven interaction graphs of the power system. By examining the interaction patterns among components, the localized areas where failures tend to occur are identified.
Another observation regarding the localization of failures can be seen in the tree structure of the power grid \cite{refe38}. It has been noted that when a non-cut set of lines fail simultaneously, the resulting failure tends to be localized within that non-cut area. On the other hand, if the failure occurs in an interconnecting line between multiple trees within the grid, it has the potential to propagate globally and impact a wider area of the system \cite{refe38}. In general, this category of problems remains relatively unexplored, with limited existing work using ML-based approaches. Few examples of such works are reviewed next.

Predicting the next $k$-failures and predicting next immediate failure (which is a special case of the next $k$-failures) in the cascade process in power systems have been studied in the literature using ML methods. For instance, the work in \cite{Naeini10} develops a general event precedence model, which builds a first-order absorbing Markov chain over the event streams and a run-time causal inference mechanism, which learns causal relationships between the events to predict $k$ failures that are most likely to occur next.
The work in \cite{else7} performs next step failure prediction in the cascades in general networks via a Bayesian belief network and a multi-attribute decision making method to perform ranking of the lines based on a scoring strategy relative to the features of the lines.
Prediction of the complete path of the cascade after the initial trigger is also studied in the literature. 
For instance, in \cite{init11}, the authors formulate the path prediction as a Markov search problem, which is built based on a large number of failure scenarios and DC power flow calculations. 
Prediction of the sequence of failures based on fault chains is proposed in \cite{b12} using time-varying graph recurrent neural network. In this work, the search for the sequence of failures is formulated as a partially observable Markov decision process and the temporal features are captured through a graph recurrent Q-learning algorithm to predict vulnerable fault chains with respect to the amount of cumulative load loss in the system. 
\vspace{-0.3cm}

\subsection{Cascade Corrective Measures}
\label{sec:sec3C}
Corrective measures play a crucial role in preventing the spread of failures and mitigating their impact on power systems. Implementing these measures during the slow precursor phase is particularly effective in enhancing the resilience of the system. The corrective measures include power dispatching and load shedding with the aim to balance the load and demand within the system and maintain a stable frequency \cite{refer89}. 
Another control action that can be employed to prevent the propagation of failures is intentional islanding. This involves isolating a specific portion of the power system to prevent further cascading effects. Intentional islanding can be triggered manually or automatically through the use of protective relays or other control mechanisms. The related research on these corrective measures is reviewed in the subsequent subsections. 

\subsubsection{Dispatching as the Corrective Measure}
Dispatching is the process of managing the output levels of power generators to meet the real-time demand for electricity \cite{basic1}. It can help preventing further damage due to cascading failures by bringing additional generation capacity online to compensate for the loss due to outages. Implementing the dispatching decisions can take several minutes; hence, it is essential to make dispatching decisions and identifying its necessary parameters quickly to prevent the situation from escalating \cite{refer83}. 
ML-based algorithms can play a key role in searching the best dispatching parameters (e.g., generator variables and set points) and profile (e.g., generator combination and schedule) with respect to specific circumstances. Some of the existing work, such as \cite{dispatch1, tox5, refer42}, have considered dispatching as an optimization problem to offer a preventive and resilience solution to cascading failures. ML-based optimization for generation dispatch is also used for generation control, and smart dispatching, as seen in the review  presented in \cite{dispatch2}. However, the application of ML in generation dispatching to address cascading failures is limited. One example is the application of an adaptive immune system reinforcement learning-based algorithm, which is presented in \cite{refer3}. Similar to the response of an immune system to destroy an antigen by an antibody, an overloading in power system is treated as an antigen and the success of the generation dispatching is treated as an antibody, based on which a reward-penalty scheme is built to select the combination of the generators for optimum power dispatching. 

\subsubsection{Load Shedding as the Corrective Measure}
Load shedding can play a critical role in stopping the propagation of failures by preventing overloading of transmission lines and generators and restoring the balance between supply and demand. Automating and optimizing the load shedding decision process and  improving the accuracy and speed of the decision-making process are examples of the problems in this category. These problems have been studied in the literature using various techniques including game-theory \cite{lseg1, lseg2}, optimization algorithms \cite{lseg3, lseg4}, and heuristic algorithms \cite{lseg6, lseg7}. Examples of ML applications to these problems are reviewed next.

In the literature, ML has been used for determining the optimal load shedding amount \cite{loadshed1, loadshed5, loadshed6}, optimal load shedding policy (i.e., sequence of load shedding actions) \cite{tox8, refer4} and the location of load shedding \cite{loadshed5} for mitigating cascading failures. 
For instance, the work in \cite{loadshed1} employs a deep neural network with an specialized loss function for predicting the required load shedding amount. The proposed network captures the non-linear and non-convex relationship of the real-time operating states and acts as a multi-input multi-output model for the risk-averse emergency load shedding (ELS) scheme. Similarly, \cite{loadshed5} uses a neural network for ELS to capture the relation between the generation and load loss, spinning reserve capacity, and sustaining frequency of the system to determine the total required load shedding amount. The work in \cite{loadshed6} proposes a GCN model for predicting the optimal load-shedding for minimizing the line overload. GCN captures the correlation between AC power flow values along with the topological information to determine the values of power dispatch and suggest appropriate amount of required load shedding. 
The work in \cite{refer4} propose a deep reinforcement learning framework to guide the control processes including generator dynamic breaking and under voltage load shedding events. The work involves modeling the power system as a Markov decision process (MDP) and using a deep neural network as the Q-function approximator to estimate the control parameters.
In \cite{tox8}, a MDP is developed to learn the optimal load shedding policy to minimize the expected cumulative cost in terms of the number of line failures and load shedding.

\newcolumntype{C}{>{\centering\arraybackslash}X} 
\setlength{\extrarowheight}{1pt} 

\begin{table*}
\caption{Classification of the references with a focus on the application of  ML in cascading failure analysis based on the proposed categories and taxonomy in Fig.~\ref{fig:fig1}.}
\label{table:table1}
    \begin{tabularx}{\textwidth}{CCCC}
        \toprule
        \textbf{Phases}	& \textbf{Applications}	& \textbf{Categories}  &  \textbf{References}\\
        \midrule
			Normal Phase & Vulnerability Analysis  & Search-based Approach & \cite{refer77}, \cite{b4}, \cite{b5}, \cite{c10}\\
            \cline{3-4} & & Classification-based Approach & \cite{refe6}, \cite{tox15} \\
            \cline{3-4} & & Regression-based Approach & \cite{refe6}, \cite{c1}, \cite{refer76} \\
			\cline{2-4} & Network Hardening & &  - \\
			\cline{2-4} & Model and Simulation & Power Flow and Failure Propagation Calculations & \cite{refer41}, \cite{else5}, \cite{b18}\\
            \cline{3-4} & & Modeling and Characterizing Interactions in Cascading Process & \cite{refe29}, \cite{tox1}, \cite{refer49}, \cite{refer86}  \\
            \cline{3-4} & & Modeling Triggers of Cascading Failure & \cite{refer33}, \cite{refer43} \\
			\hline
			Cascade Precursor Phase & Initial Fault Analysis & &  \cite{v2}, \cite{ref4}, \cite{N12}, \cite{N13}, \cite{N14}, \cite{N15}, \cite{refer74}, \cite{refer38}, \cite{refer87} \\		
            \cline{2-4} & Cascade Prediction & Prediction of Cascade Size & \cite{refe13}, \cite{igraph2}, \cite{refe34}\\
			\cline{3-4} & & Prediction of Cascade Evolution and Temporal Aspects & \cite{m38}, \cite{p5} \\
			\cline{3-4} & & Prediction of Cascade Path and Areas at Risk & \cite{Naeini10}, \cite{else7}, \cite{init11}, \cite{b12} \\
            \cline{2-4} & Cascade Corrective Measures & Dispatching as Cascading Corrective Measure & \cite{refer3}\\
            \cline{3-4} & & Load Shedding as Cascading Corrective Measure & \cite{refer4}, \cite{loadshed1}, \cite{loadshed5}, \cite{loadshed6}, \cite{tox8}  \\
            \cline{3-4} & & Islanding as Cascading Corrective Measure & \cite{tox16}, \cite{island2} \\
			\hline
			Cascade Escalation and Termination Phase & & & - \\
			\hline
			Cascade Restoration Phase	& Failure Recovery & & \cite{init3}, \cite{rest5}, \cite{rest6} \\
			\cline{2-4} & Root Cause Analysis &  & - \\
			\bottomrule
    \end{tabularx}
\end{table*}

\subsubsection{Controlled Islanding as the Corrective Measure}
Controlled islanding is the process of dividing the power system into smaller, independent sections, which can continue operating independently using local resources. Such islanding can help mitigating cascading failures by containing the effects of the disturbance to a localized area, and preventing the spread to other parts of the system. 
Controlled islanding to prevent cascading failures has been studied in the literature in different forms including as an optimization problem \cite{else4, tox7} and graph-theoretic problem \cite{else9, tox11}. 
 
ML techniques have also been applied to the problem of controlled islanding to improve the decision making process and speeding it up instead of relying on time-consuming calculations of large number of real-time variables.
For instance, \cite{tox16} suggests a clustering algorithm combined with non-linear programming to identify groups of coherent generators. The clustered generators are considered as the core of the islands and the nearby generators and loads are identified to construct the sub-network of the islands using Djikstra algorithm. 
The work in \cite{island2} proposes a deep GCN for graph partitioning to mitigate the load-generation imbalance within the islands. A specialized loss function is designed to cluster each bus into an island based on the generator coherency and to assign independent components in different islands depending on their distances in the topology. 

\section{ML-based Cascade Analysis in Escalation Phase}
\label{sec:sec4}
The escalation phase in the cascade process of power systems is characterized by the rapid propagation of failures throughout the system. The rapid spread of failures during this phase limits the available options for intervention and poses significant challenges for system operators. Due to the severity and complexity of the escalation phase, there is a scarcity of studies specifically focused on addressing this particular phase. However, the importance of studying and addressing the escalation phase should not be underestimated. Despite the challenges, gaining insights into the dynamics and behavior of cascades during this phase is crucial for enhancing the resilience and robustness of power systems. Exploring techniques and methodologies to effectively intervene and mitigate the rapid spread of failures can contribute to reducing the severity and duration of blackouts, and minimizing economic losses.
Further research and attention are needed to delve into the complexities of the escalation phase. By understanding the underlying mechanisms and identifying potential measures to halt or slow down the cascade's progression, operators can improve their response capabilities and enhance the overall resilience of power systems.
\vspace{-0.1cm}
\section{ML-based Cascade Analysis in Post Termination Phase}
\label{sec:sec5}
The focus of this category is on addressing the challenges that arise after the cascade process has concluded and a blackout has already occurred. The primary focus of these studies is on cascade fault recovery, service restoration, and conducting root cause analysis of the cascade event. Cascade fault recovery and service restoration involve various critical tasks aimed at restoring power and ensuring the resumption of normal operations. 
These tasks include assessing the extent of the cascade and identifying the root causes of the failures, restoring the power through black-starting of the generators using auxiliary power sources, repairing or replacing damaged components such as transmission lines and transformers, and synchronizing the restored sections of the system \cite{rec1}. Determining the most effective sequence of actions for restoration is crucial, considering the urgency of restoring power as quickly as possible. The utilization of ML techniques to support post-cascade operations is still relatively limited. In this section, the existing work after the cascade termination has been reviewed under two categories of failure recovery process and root-cause analysis.

\vspace{-0.3cm}
\subsection{Cascade Failure Recovery}
\label{sec:sec5A}
In the cascade recovery process, one of the key challenges is determining the optimal order for restoring the failed components. Restoring the failed and damaged components in the right sequence is crucial to ensure a smooth and efficient recovery of the power system.
Identifying the order of restoration involves considering various factors, such as the criticality of the component, the availability of resources and personnel, time required for repairs, the dependencies between components, and the overall system stability. 
This problem has been studied in the literature through optimization-based approach \cite{refer42} and heuristic approach \cite{refer36}. ML techniques can also be employed to address this problem by analyzing historical data, system topology, and real-time information to identify patterns and dependencies among the failed components and to predict the optimal order of component restoration. For instance, the works in \cite{init3, rest5, rest6} find the bus/branch recovery sequences after a cascading failure using a Q-learning mechanism, to achieve topological restoration with the minimum number of repairing steps. The purpose of the Q-learning is to obtain maximum reward (restoration) with minimum component recovery.

\vspace{-0.3cm}
\subsection{Cascade Root Cause Analysis}
\label{sec:sec5B}
Root cause analysis (RCA) is an important process, which enables understanding the causalities in the system and supports fault mitigation, system upgrade, and investment decisions. The ML-based approaches to RCA can allow tracing the potential chain of cascade to identify the causes, triggering, and fueling factors by extracting and inferring causality and interaction information from large, complex and incomplete cascade and system data. 
RCA approaches has been applied to power systems \cite{root1, RCA-Power2,RCA-Power3,RCA-Power4,RCA-Power5} and other networked systems, such as wireless networks \cite{RCA3, RCA4}.
Many RCA approaches are based on statistical analysis to infer association among variables and  identify dependencies among time-series \cite{RCA-TimeSeries1,RCA-TimeSeries2}. Probabilistic graphical models \cite{RCA-GM2, RCA-GM4} have also been adopted for RCA; however, they require specifying conditional dependency structure among variables. Recently, various ML models have been used to enable inferring the dependencies and interactions in data for RCA \cite{RCA2}. In power systems, many of the RCA approaches are focused on fault cause identification. The examples include RCA for power transformers fault \cite{RCA-Power2}, transmission line fault \cite{RCA-Power3, RCA-Power4}, and fault cause identification using waveform measurements in distribution networks \cite{RCA-Power5}. These works mainly perform individual waveform or time-series analyses for RCA to classify the causes of faults and does not consider interactions or dependencies among various components and their failures. RCA has also been applied to cascading failures problem in interdependent power and communication systems with an algorithmic approach in \cite{root1}. In the latter work, the node measurements are not considered and instead the RCA is carried out over interdependency relations in the form of Boolean Logic relations, which are assumed to be known. The application of ML to RCA in power systems with regards to cascading failures is still very limited and yet to be explored. 

\section{Conclusion}
\label{sec:sec6}
Cascading failures present a significant threat to power grids, necessitating extensive research efforts in their analysis and understanding.
The combination of advancements in monitoring technologies, the availability of vast amounts of power system data, and the emergence of intelligent algorithms have made ML techniques increasingly appealing for analyzing cascading failures.
The presented review in this survey provides a comprehensive overview of ML-based techniques employed in the analysis of cascading failures in power systems. By categorizing these techniques based on the different phases of the cascade process and examining research on cascade resiliency prior to and after the occurrence of cascades, this survey offers new insights and a systematic understanding of ML's role in modeling, analyzing, and mitigating cascading failures. The organization and presentation of these works into relevant categories contribute to a better understanding of the strengths and limitations of ML approaches in addressing cascading failures. The gaps in the existing research also show the importance of further research in this domain to fully exploit the potential of ML in cascade analysis. As power systems continue to evolve and face new challenges, for instance, due to stochastic and uncertain renewable resources and incorporation of energy storage systems, the integration of ML techniques holds great potential for advancing our understanding of cascading failures and improving preventive and corrective measures under new circumstances.
\vspace{-0.2cm}

\section*{Acknowledgement}
This material is based upon work supported by the National Science Foundation under Grants No. 2238658 and No. 2118510.  Any opinions, findings, and conclusions or recommendations expressed in this material are those of the author(s) and do not necessarily reflect the views of the National Science Foundation.

\bibliography{arXiv.bib}{}
\bibliographystyle{IEEEtran}
\end{document}